# Threshold Current for Field-free Switching of the In-plane Magnetization in the Three-terminal Magnetic Tunnel Junction

Hongjie Ye, and Zhaohao Wang

***Abstract*: Three-terminal magnetic tunnel junction (MTJ), where non-volatile magnetization state can be switched via spin orbit torque (SOT), is attracting massive research interests since it is featured by high speed, low power, nearly unlimited endurance, etc. The threshold switching current is a key parameter for MTJ as it determines the energy efficiency. Here, with the Routh-Hurwitz criterion, we theoretically derive the threshold current for switching in-plane magnetization in the three-terminal MTJ. Two devices with field-free switching mode are investigated. The one is the Type-x device switched by the combination of SOT and spin transfer torque (STT). The other is the three-terminal MTJ with a canted easy-axis. To the best of our knowledge, this is the first theoretical work on the threshold switching current for these two devices. Our developed theoretical method shows clear physical picture, meanwhile good agreement between theoretical derivation and numerical simulation is achieved.**

## I. Introduction

MAGNETIC random access memory (MRAM) is considered as the promising candidate for the future universal memory thanks to the various advantages. The switching mechanism of MRAM has long been a research focus since it directly influences the write performance. Currently, three-terminal magnetic tunnel junction (MTJ) which can be switched by the spin orbit torque (SOT), is becoming popular for the next-generation MRAM (generally called SOT-MRAM), owing to ultralow power, sub-nanosecond switching speed and high reliability [1]. Therefore, plenty of theoretical study on the three-terminal MTJ has been carried out. Among them, the threshold switching current is paid more attention because it is of importance to evaluate the energy efficiency. However, most of them is focused on the perpendicular magnetization [2]-[7], whereas in-plane magnetization has been studied by only a few literatures [8]. In reality, the reported SOT-MRAM chips and arrays are fabricated with in-plane-magnetized three-terminal MTJs [9]-[11]. Therefore more work is needed for studying the threshold switching current of the in-plane-magnetized three-terminal MTJ.

For practical application, the SOT-MRAM must be switched in the absence of the external magnetic field, i.e. field-free deterministic switching. According to classification criterion of [12], the in-plane-magnetized three-terminal MTJ is divided into two categories: Type-x and Type-y devices. The Type-y device could be deterministically switched in the field-free manner. Conversely, the deterministic switching of Type-x device requires the assistance of a magnetic field. For avoiding the use of magnetic field, two solutions have been proposed and experimentally validated. The one is to combine the SOT and spin transfer torque (STT) [13], as shown in Fig. 1(a). The other is to cant the easy-axis of the MTJ for breaking the symmetry [14], as illustrated in Fig. 1(b). To the best of our knowledge, no related theoretical work has been carried out for solving the threshold switching current of these two field-free Type-x devices. In this paper, we develop a novel theoretical method to resolve this problem, providing insight into the characteristic of magnetization switching.

The remainder of this paper is organized as follows. The theoretical models and detailed derivation are described in section II. Then, the results are shown and discussed in section III. Finally, we conclude this work in section IV.

## II. Theoretical derivation

First, we theoretically study the Type-x device which is switched by the combination of SOT and STT. Its detailed structure is shown in Fig. 1(a). Two currents are applied to the SOT channel and MTJ, respectively, for inducing the SOT and STT. The deterministic switching of magnetization in the free layer is achieved under the joint effect of SOT and STT [13]. The magnetization dynamics can be modelled by the following modified Landau-Lifshitz-Gilbert (LLG) equation.

$$\frac{\partial \vec{m}}{\partial t} = -\gamma \vec{m} \times \vec{H}_{eff} + \alpha \vec{m} \times \frac{\partial \vec{m}}{\partial t} \\ - \gamma H_{SOT} \vec{m} \times (\vec{\sigma} \times \vec{m}) - \gamma H_{STT} \vec{m} \times (\vec{m}_p \times \vec{m}) \quad (1)$$

This work was supported by the National Natural Science Foundation of China under Grant 62171013, the National Key Research and Development Program of China (Nos. 2021YFB3601303, 2021YFB3601304, 2021YFB3601300), and the Fundamental Research Funds for the Central Universities. (Corresponding author: Zhaohao Wang).

H. Ye and Z. Wang are with the Fert Beijing Institute, MIIT Key Laboratory of Spintronics, School of Integrated Circuit Science and Engineering, Beihang University, Beijing 100191, China. (e-mail: zhaohao.wang@buaa.edu.cn).

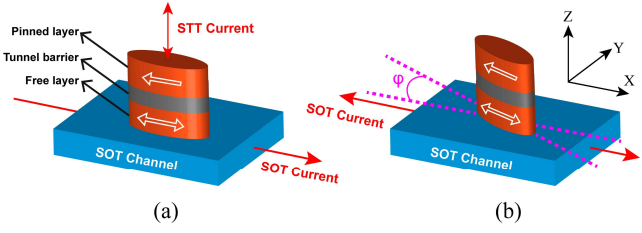

Fig. 1. Structure of the studied device. (a) Type-x device switched by the combination of SOT and STT. (b) Type-x device with the canted easy-axis.

where $\gamma$ and $\alpha$ are the gyromagnetic ratio and the Gilbert damping constant, respectively. $\vec{\sigma}$ is the unit vector of SOT-induced spin polarization. $\vec{m}_p$ is the unit vector of magnetization in the pinned layer. $\vec{\sigma}$ and $\vec{m}_p$ are alined to y and x-axis, respectively. $\vec{H}_{eff}$ is the effective field, as

$$\vec{H}_{eff} = (H_k m_x, 0, -4\pi M_s m_z) \qquad (2)$$

where $H_k$ is the effective in-plane anisotropy field, $M_s$ is the saturation magnetization, $m_x$ and $m_z$ are the x- and z-components of the unit magnetization vector, respectively. $H_{SOT}$ and $H_{STT}$ are effective fields induced by SOT and STT, respectively, as

$$H_{SOT} = \frac{\hbar \theta_{SHE} J_{SOT}}{2 e t_F M_s} \qquad (3)$$

$$H_{STT} = \frac{\hbar P J_{STT}}{2 e t_F M_s} \qquad (4)$$

where $\hbar$ is the reduced Planck constant, $\theta_{SHE}$ is the spin Hall angle, $P$ is the spin polarization efficiency of the STT effect, $e$ is the electron charge, $t_F$ is the free layer thickness, $J_{SOT}$ and $J_{STT}$ are the current densities of SOT and STT, respectively.

There are two equilibria in (1), which is given by substituting $\partial m_{x,y,z}/\partial t = 0$ into (1), as

$$4\pi M_s m_y m_z + H_{SOT} m_x m_y - H_{STT}\left(m_y^2 + m_z^2\right) = 0 \qquad (5)$$

$$-(4\pi M_s + H_k) m_x m_z - H_{SOT}(m_x^2 + m_z^2) + H_{STT} m_x m_y = 0 \qquad (6)$$

$$H_k m_x m_y + H_{SOT} m_y m_z + H_{STT} m_x m_z = 0 \qquad (7)$$

The two equilibria are near the point $(1,0,0)$ and $(-1,0,0)$. Assume that the initial equilibrium is $(1,0,0)$, then $m_{y,z} \approx 0$. In addition, generally $4\pi M_s$ is far larger than $H_{STT}$. Thus $H_{STT}(m_y^2 + m_z^2)$ can be omitted in (5), as

$$4\pi M_s m_y m_z + H_{SOT} m_x m_y = 0 \qquad (8)$$

Combining (7) and (8), two equilibria are obtained, as

$$E = \pm \left( m_x, \frac{H_{STT} H_{SOT} m_x}{4\pi M_s H_k - H_{SOT}^2}, -\frac{H_{SOT} m_x}{4\pi M_s} \right) \qquad (9)$$

The characteristic polynomial of the Jacobian matrix of function $\partial m_{x,y,z}/\partial t = f_{x,y,z}(m_x, m_y, m_z)$ is written as

$$P = a_0 \lambda^3 + a_1 \lambda^2 + a_2 \lambda + a_3 \qquad (10)$$

According to Routh-Hurwitz criterion, the condition making an equilibrium stable is $\Delta_1, \Delta_2 > 0$ when $a_0 > 0$, or $\Delta_1, \Delta_2 < 0$ when $a_0 < 0$. $\Delta_1$ and $\Delta_2$ are given by

$$\Delta_1 = a_1 \qquad (11)$$

$$\Delta_2 = \begin{Vmatrix} a_1 & a_0 \\ a_3 & a_2 \end{Vmatrix} \qquad (12)$$

It is easily inferred from Jacobian matrix that $a_0 < 0$ in this case. Thus the equilibrium can be unstable under the condition of $\Delta_1 > 0$. Substituting the equilibria (9) into $\Delta_1$ and considering $m_x \approx 1$, the switching condition can be derived as

$$H_{STT} > \frac{\alpha}{4\pi M_s H_k}\left(H_k + 2\pi M_s - \frac{3 H_{SOT}^2}{8\pi M_s}\right) \\ \times \left(4\pi M_s H_k - H_{SOT}^2\right) \qquad (13)$$

Substituting the (4) into (13), the threshold switching current density can be solved, as

$$J_{STT,th} = \frac{2\alpha e t_F M_s}{\hbar P}\left(H_k + 2\pi M_s - \frac{3 H_{SOT}^2}{8\pi M_s}\right) \\ \times \left(1 - \frac{H_{SOT}^2}{4\pi M_s H_k}\right) \qquad (14)$$

When $H_{SOT} = 0$, the equation (14) is consistent with the expression of threshold current density for pure STT switching [15]. When $J_{STT,th} = 0$, the equation (14) could give $H_{SOT} = \sqrt{4\pi M_s H_k}$ which is identical to the critical condition for activating the instability of a pure Type-x device [8]. Thus equation (14) is compatible with the reported conclusion.

Following the above methodology, we can calculate the threshold switching current density for the canted Type-x device. The corresponding structure is shown in Fig. 1(b), where the easy axis of the MTJ is canted with respect to the SOT channel. In this scenario, the LLG equation is modified as

$$\frac{\partial \vec{m}}{\partial t} = -\gamma \vec{m} \times \vec{H}_{eff} + \alpha \vec{m} \times \frac{\partial \vec{m}}{\partial t} - \gamma H_{SOT} \vec{m} \times (\vec{\sigma} \times \vec{m}) \qquad (15)$$

As the applied SOT current is canted from the easy-axis, the unit vector of spin polarization $\vec{\sigma}$ can be expressed as $(\sin\varphi, \cos\varphi, 0)$, where $\varphi$ is the canted angle.

Using the above approach and considering $m_x \approx 1, m_z \ll 1$, we can obtain the threshold switching current density for the canted Type-x device, as

$$J_{SOT,th} = \frac{e t_F M_s}{\hbar \theta_{SHE}} \cdot \frac{4\pi M_s H_k \sin\varphi}{\alpha(H_k + 2\pi M_s)\cos^2\varphi} \\ \times \left[\sqrt{1 + \frac{\alpha^2(H_k + 2\pi M_s)^2 \cos^2\varphi}{\pi M_s H_k \sin^2\varphi}} - 1\right] \qquad (16)$$

In the cases of $\varphi \to 0$ and $\varphi \to \pi/2$, equation (16) is consistent with the formula of a pure Type-x device [8] and a

Type-y device [16], respectively. If $\tan\varphi \gg \alpha(H_k + 2\pi M_s)/\sqrt{\pi M_s H_k}$, equation (16) can be approximately expressed as

$$J_{SOT,th} \approx \frac{2\alpha e t_F M_s (H_k + 2\pi M_s)}{\hbar \theta_{SHE} \sin\varphi} \quad (17)$$

which has the similar form to the threshold switching current density of the canted Type-z device [17].

### III. RESULTS AND DISCUSSIONS

To validate the above theoretical formulas, we numerically solved the LLG equation with the fourth-order Runge-Kutta method for the comparison. The main parameters are configured as follows [12]-[16], unless otherwise stated. $\alpha = 0.02$, $t_F = 1.5$ nm, $M_s = 1000$ emu/cm$^3$, $\theta_{SHE} = 0.3$, $P = 0.6$, $H_k = 300$ Oe, which could reflect the typical fabrication process of the in-plane magnetized three-terminal MTJ. The comparison between theoretical formulas and numerical simulation is shown in Fig. 2.

Figure 2(a) shows the $J_{STT,th}$ as a function of the applied $J_{SOT}$, corresponding to the device of Fig. 1(a). It is seen that the $J_{STT,th}$ is reduced by the increasing $J_{SOT}$, consistent with the experimental report [13]. Moreover, good agreement between the theoretical formula and numerical results is achieved. This also demonstrates that the approximation condition used in our theoretical derivation is reasonable. Figure 2(b) shows the dependence of $J_{STT,th}$ on the Gilbert damping constant ($\alpha$). A clear linearity can be seen from both theoretical formula and numerical simulation, also showing good agreement between them. Figure 2(c)-(d) corresponds to the device of Fig. 1(b). As can be seen from Fig. 2(c), the threshold switching current decreases as the canted angle increases. Moreover, $J_{SOT,th}$ is almost proportional to $1/\sin\varphi$, but the deviation occurs in the case of small $\varphi$. Not only does this tendency validate the accuracy of (16)-(17), but also it is highly consistent with the reported experimental results [18]-[19]. In addition, Fig. 2(d) demonstrates that $H_k$ has little influence on the $J_{SOT,th}$, which can be explained by the fact of $2\pi M_s \gg H_k$.

### IV. CONCLUSION

We have theoretically derived the threshold current for the field-free switching of Type-x MTJs. The accuracy of the results was verified by good agreement between theoretical formulas and numerical simulation. The derived theoretical formulas reveal the influence of key parameters on the magnetization switching. Moreover our developed theoretical method is featured by conciseness and clarity. Our work will guide the design and optimization of the Type-x MTJs.

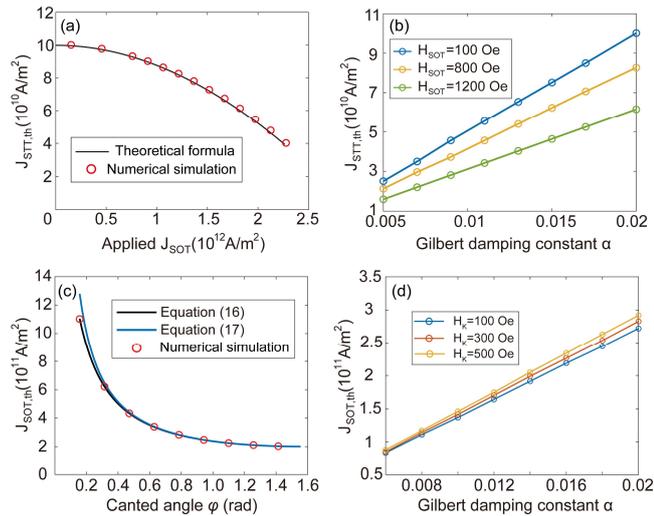

Fig. 2. Good agreement between theoretical formulas and numerical simulation. (a) and (b) are results of the Type-x device switched by the combination of SOT and STT. (c) and (d) are results of the Type-x device with the canted easy-axis. (a) $J_{STT,th}$ as a function of the applied $J_{SOT}$. (b) The dependence of $J_{STT,th}$ on the Gilbert dampiing constant. (c) $J_{SOT,th}$ as a function of the canted angle. (d) The dependence of $J_{SOT,th}$ on the Gilbert damping constant.

### REFERENCES

[1] Z. Guo, J. Yin, Y. Bai, D. Zhu, K. Shi, G. Wang, K. Cao, and W. Zhao, "Spintronics for energy-efficient computing: An overview and outlook," *Proc. IEEE*, vol. 109, no. 8, pp. 1398–1417, Aug. 2021, doi: 10.1109/JPROC.2021.3084997.

[2] K.-S. Lee, S.-W. Lee, B.-C. Min, and K.-J. Lee, "Threshold current for switching of a perpendicular magnetic layer induced by spin Hall effect," *Appl. Phys. Lett.*, vol. 102, no. 11, p. 112410, Mar. 2013, doi: 10.1063/1.4798288.

[3] T. Taniguchi, "Theoretical condition for switching the magnetization in a perpendicularly magnetized ferromagnet via the spin Hall effect," *Phys. Rev. B*, vol. 100, no. 17, p. 174419, Nov. 2019, doi: 10.1103/PhysRevB.100.174419.

[4] D. Zhu and W. Zhao, "Threshold Current Density for Perpendicular Magnetization Switching Through Spin-Orbit Torque," *Phys. Rev. Appl.*, vol. 13, no. 4, p. 044078, Apr. 2020, doi: 10.1103/PhysRevApplied.13.044078.

[5] S. Wasef and H. Fariborzi, "Theoretical Study of Field-Free Switching in PMA-MTJ Using Combined Injection of STT and SOT Currents," *Micromachines*, vol. 12, no. 11, p. 1345, Oct. 2021, doi: 10.3390/mi12111345.

[6] D. H. Kang, J.-H. Byun, and M. Shin, "Critical switching current of a perpendicular magnetic tunnel junction owing to the interplay of spin-transfer torque and spin-orbit torque," *J. Phys. D: Appl. Phys.*, vol. 54, no. 43, p. 5001, Aug. 2021, doi: 2021 10.1088/1361-6463/ac181a.

[7] H. Lim, S. Lee, and H. Shin, "Unified Analytical Model for Switching Behavior of Magnetic Tunnel Junction," *IEEE Electron Device Lett.*, vol. 35, no. 2, pp. 193-195, Feb. 2014, doi: 10.1109/LED.2013.2293598.

[8] T. Taniguchi, "Switching induced by spin Hall effect in an in-plane magnetized ferromagnet with the easy axis parallel to the current," *Phys. Rev. B*, vol. 102, no. 10, p. 104435, Sep. 2020, doi: 10.1103/PhysRevB.102.104435.

[9] M. Natsui, A. Tamakoshi, H. Honjo, T. Watanabe, T. Nasuno, C. Zhang, T. Tanigawa, H. Inoue, M. Niwa, T. Yoshiduka, Y. Noguchi, M. Yasuhira, Y. Ma, H. Shen, S. Fukami, H. Sato, S. Ikeda, H. Ohno, T. Endoh, and T. Hanyu, "Dual-Port SOT-MRAM Achieving 90-MHz Read and 60-MHz Write Operations Under Field-Assistance-Free Condition," *IEEE J. Solid-State Circuits*, vol. 56, no. 4, pp. 1116–1128, Apr. 2021, doi: 10.1109/JSSC.2020.3039800.

[10] G.-L. Chen, I-J. Wang, P.-S. Yeh, S.-H. Li, S.-Y. Yang, Y.-C. Hsin, H.-T. Wu, H.-M. Hsiao, Y.-J. Chang, K.-M. Chen, SK Z. Rahaman, H.-H. Lee, Y.-H. Su, F.-M. Chen, J.-H. Wei, S.-S. Sheu, C.-I. Wu, and D. Tang, "An 8kb spin-orbit-torque magnetic random-access memory," in *2021 International Symposium on VLSI Technology, Systems and Applications (VLSI-TSA)*, Hsinchu, Taiwan: IEEE, Apr. 2021, pp. 1–2, doi: 10.1109/VLSI-TSA51926.2021.9440096.

[11] M. Y. Song, C. M. Lee, S. Y. Yang, G. L. Chen, K. M. Chen, I J. Wang, Y. C. Hsin, K. T. Chang, C. F. Hsu, S. H. Li, J. H. Wei, T. Y. Lee, M. F. Chang, X. Y. Bao, C. H. Diaz, and S. J. Lin, "High speed (1ns) and low voltage (1.5V) demonstration of 8Kb SOT-MRAM array," in *2022 IEEE Symposium on VLSI Technology and Circuits (VLSI Technology and*

*Circuits)*, Honolulu, HI, USA: IEEE, Jun. 2022, pp. 377–378, doi: 10.1109/VLSITechnologyandCir46769.2022.9830149.

[12] S. Fukami, T. Anekawa, C. Zhang, and H. Ohno, "A spin–orbit torque switching scheme with collinear magnetic easy axis and current configuration", *Nature Nanotech.*, vol. 11, no. 7, pp. 621–625, Jul. 2016, doi: 10.1038/nnano.2016.29.

[13] C. Zhang, Y. Takeuchi, S. Fukami, and H. Ohno, "Field-free and sub-ns magnetization switching of magnetic tunnel junctions by combining spin-transfer torque and spin–orbit torque," *Appl. Phys. Lett.*, vol. 118, no. 9, p. 092406, Mar. 2021, doi: 10.1063/5.0039061.

[14] H. Honjo, T. V. A. Nguyen, T. Watanabe, T. Nasuno, C. Zhang, T. Tanigawa, S. Miura, H. Inoue, M. Niwa, T. Yoshiduka, Y. Noguchi, M. Yasuhira, A. Tamakoshi, M. Natsui, Y. Ma, H. Koike, Y. Takahashi, K. Furuya, H. Shen, S. Fukami, H. Sato, S. Ikeda, T. Hanyu, H. Ohno, and T. Endoh, "First demonstration of field-free SOT-MRAM with 0.35 ns write speed and 70 thermal stability under 400°C thermal tolerance by canted SOT structure and its advanced patterning/SOT channel technology," in *2019 IEEE International Electron Devices Meeting (IEDM)*, San Francisco, CA, USA: IEEE, Dec. 2019, p. 28.5.1-28.5.4, doi: 10.1109/IEDM19573.2019.8993443.

[15] Z. Diao, Z. Li, S. Wang, Y. Ding, A. Panchula, E. Chen, L.-C. Wang, and Y. Huai, "Spin-transfer torque switching in magnetic tunnel junctions and spin-transfer torque random access memory," *J. Phys.: Condens. Matter*, vol. 19, no. 16, p. 165209, Apr. 2007, doi: 10.1088/0953-8984/19/16/165209.

[16] L. Liu, C.-F. Pai, Y. Li, H. W. Tseng, D. C. Ralph, and R. A. Buhrman, "Spin-Torque Switching with the Giant Spin Hall Effect of Tantalum," *Science*, vol. 336, no. 6081, pp. 555–558, May 2012, doi: 10.1126/science.1218197.

[17] D.-K. Lee and K.-J. Lee, "Spin-orbit Torque Switching of Perpendicular Magnetization in Ferromagnetic Trilayers," *Sci. Rep.*, vol. 10, no. 1, p. 1772, Feb. 2020, doi: 10.1038/s41598-020-58669-1.

[18] Y. Takahashi, Y. Takeuchi, C. Zhang, B. Jinnai, S. Fukami and H. Ohno, "Spin-orbit torque-induced switching of in-plane magnetized elliptic nanodot arrays with various easy-axis directions measured by differential planar Hall resistance," *Appl. Phys. Lett.*, vol. 114, no. 1, p. 012410, Jan. 2019, doi: 10.1063/1.5075542.

[19] Y.-T. Liu, Y.-H. Huang, C.-C. Huang, Y.-C. Li, C.-L. Cheng, and C.-F. Pai, "Field-Free Type-x Spin-Orbit-Torque Switching by Easy-Axis Engineering," *Phys. Rev. Appl.*, vol. 18, no. 3, p. 034019, Sep. 2022, doi: 10.1103/PhysRevApplied.18.034019